\documentclass[%
 aip,
cp,  
 amsmath,amssymb,
 reprint,%
]{revtex4-2}

\usepackage{graphicx}
\usepackage{ulem}
\usepackage{dcolumn}
\usepackage{bm}

\usepackage{xcolor}
\usepackage[utf8]{inputenc}
\usepackage[T1]{fontenc}
\usepackage{mathptmx} 

\begin{document}

\title{How Stellar Stream Torsion may reveal aspherical Dark Matter Haloes
}

\author{Adriana Bariego-Quintana} 
 \email[Corresponding author: ]{adriana.bariego@gmail.com}
 \affiliation{
  IFIC-Univ. Valencia, c/ Catedrático José Beltrán, 2, E-46980 Paterna, Valencia, Spain.
}

\date{\today} 

\begin{abstract}
Flat rotation curves v(r) naturally follow from elongated (prolate) Dark Matter distributions, as shown by our earlier competitive fits to the SPARC database. Intending to probe that distortion of the DM halo one needs observables not contained by the galactic plane.
Stellar streams are caused by tidal stretching of massive substructures such as satellite dwarf galaxies, and would lie on a plane should the DM-halo gravitational field be spherically symmetric. But if the field does not display such spherical symmetry, stellar trajectories, as well as stellar streams, should torsion out of the plane. 
This is where the torsion of the stream can be of use: it is a local observable that measures the deviation from planarity of a curve; thus, it quantifies how noncentral the gravitational potential is.
We have performed small simulations to confirm that indeed a galactic central force  produces negligible torsion, and quantified the torsion for prolate haloes instead. Examining observational data, we select several streams at large distance from the galactic center, as most promising for the study, and by means of helicoidal fits extract their differential torsion.
We see  that their torsion is much larger than expected for a central spherical bulb alone, pointing to an elongated Milky Way halo. 
\end{abstract}
\maketitle

\section{\label{sec:level1} Introduction}
There is substantial observational evidence for a Dark Matter (DM) component at different scales in the Universe; from rotation curves to the large scale structure we identify a non-negligible quantity of mass that we are missing in ordinary observations. In this work we focus on the observation of DM at galactic scale, in particular, we study a topic related to the rotation curve problem. Orbital equilibrium outside of a spherical source predicts a decrease of the velocity with which a star orbits around a galaxy as it moves away from its centre, $v(r) \propto r^{-1/2}$.  But on the contrary, empirical rotation curves seem to flatten out for a lot of galaxies $v(r)\propto C$ \cite{1978ApJ...225L.107R}. 

There has a been great deal of attempts to solve this problem. An alternative approach to DM is to modify the basic laws of mechanics with the Modified Newtonian Dynamics, also called MOND; however, these models run into problems at larger cosmological scales. If DM is accepted, an alternative to an isothermal spherical distribution is to accept a shape modification of the DM gravity source.
In the 3-dimensional cosmos that we live in we would expect rotation curves to decrease with distance to the galactic center $v^2\propto 1/r$, in a similar way in a 2-dimensional cosmos we would see constant velocities $v^2\propto C$. Current observations show that the velocities in rotation curves are constant $v^2\propto C$. The question is how to achieve that dimensional reduction living in a 3-dimensional cosmos.

A spherical DM distribution has to be fine-tuned with an isothermal density profile $ \rho(r)\sim r^{-2}$  to explain the flatness of the rotation curves. Whereas the extreme case of a cylindrical halo of linear mass density $\lambda$ naturally explains the flatness $v = \sqrt{2G\lambda}$ \cite{2021Univ....7..346L}. Since the rotation curve is measured to a finite radius the source does not need to be infinitely cylindrical: it is sufficient that it is an elongated DM halo \cite{PhysRevD.107.083524}. One additional issue that arises concerns the distinction between spherical haloes with an isothermal profile and elongated haloes with arbitrary profiles using as objects of study the galactic rotation curves that are observable within the galactic plane. This differentiation poses a challenge due to insufficient data points outside of the galactic disk. Therefore, the inclusion of out-of-plane observables could become advantageous, for this purpose stellar streams emerge as promising candidates.
\section{\label{sec:level3} N-Body simulations of stellar streams and fitting}

We simulate clusters of $N=100$ to $1000$ point-like stars initially located at randomly distributed positions over a sphere of radius $R_0=2$ kpc located at a distance of $r_0 = 10$ to $30$ kpc from the galactic center. The simulated stars are given random masses taking values in the range $m_\star = (1,20)$ $M_\odot$, and have a common initial velocity in the y direction $v_{\star,y} \sim 220 $ kpc/s to simulate the movement around the gravitational source. Some stars in the cluster receive a random initial velocity $\Delta \textbf{v} = G m_{cluster}/2r_0$ to add some dispersion to the cloud of point-like particles. 
After declaring the initial conditions of the particles in the cluster (positions and velocities), we make it evolve around a gravitational source from an initial time $t=0$ My to different final times. 

\textbf{ IMPLEMENTATION OF THE N-BODY SIMULATIONS}\\

Let us first assume we have $N$ particles $i=1,2,...,N$ and each of them feels the gravitational attraction of all the others following Newton's Law of Universal Gravitation 
\begin{eqnarray}
    \textbf{a}_i = - G\sum^{N-1}_{j\neq i}  m_j \frac{\textbf{r}_j- \textbf{r}_i}{|\textbf{r}_j- \textbf{r}_i|^3},
\end{eqnarray}
taking $G=4.53\cdot10^{-12}$ $ \text{kpc}^3\text{My}^{-2}  M_\odot^{-1}$ as the gravitational constant, $m_j$ being the mass of the $j=1,2,...N-1$ of the remaining stars and $\textbf{r}_j- \textbf{r}_i$ the distance between the point-like stars.
The next step is to allow this cluster of particles to evolve under the influence of an external force generated by a second gravitational source in whose symmetry center we locate our coordinate origin. In this work we want to consider different kinds of gravitational sources: \\

\textbf{Spherical gravitational source}, we have the possibility of simulating a spherical galactic bulge or a DM spherical halo. The only difference between both possibilities in this context is the total mass of the central source, dark matter sources are thought to have masses in the range $M\in (10^{10}, 10^{12})$ $M_\odot$ whereas galactic sources are  found to be slightly less massive $M\in (10^{9}, 10^{11})$ $M_\odot$. The gravitational force felt by the individual stars in the stream due to this spherical source is then $ \textbf{a}_i = - GM\sum_i^{N}  \frac{ \textbf{r}_i}{|\textbf{r}_i|^3}.$ \\

\textbf{Cylindrical gravitational source}, to simulate a DM cylinder toy-model. We consider a cylinder of linear mass density $\lambda=M/L$ with a value obtained from galactic rotation curves $v_{rot} = \sqrt{2G\lambda}$. There will be a difference between the vertical and horizontal forces felt by the stars in the stream $\textbf{a}_i = - \frac{2G\lambda}{x^2+y^2}  (x,y,0).$

After specifying the initial conditions of the particles we can describe their movement in time due to their binding forces and to the external forces created by the galactic/DM gravitational sources following Euler's method. It updates the position and velocity of the particles in a certain time step $\Delta t = t_f / N_{it}$, $t_f$ being the final time and $N_{it}$ the number of iterations and $\textbf{f}$ denoting the acceleration: $\textbf{v}_{i+1} = \textbf{v}_i +\frac{\Delta t}{2} \cdot \textbf{f }(\textbf{r }_i+\frac{\Delta t}{2} \textbf{v}_i)$ and $\textbf{r}_{i+1} = \textbf{r}_i + \Delta t \cdot \textbf{v}_i$.\\



\textbf{  FITTING OF THE N-BODY STREAMS }\\

Once the N-body simulations of the streams for chosen final times, $t_f$, have been arranged, we will fit the stream numerical data to parametric curves in cartesian coordinates following a simple functional dependence $\textbf{r}(t) = \textbf{r}_0 + (A \cdot\cos(\omega\cdot t +\phi), B \cdot\sin(\omega \cdot t + \phi), C\cdot t),$ where $\{$A, B, C, $\omega$, $\phi$, $\textbf{r}_0 \}$ is the fitting parameters set such that A and B describe the elliptic projection on the XY plane starting with an angular shift $\phi$, C describes the advance of the helix in the Z direction and $\omega$ is the angular velocity in the XY plane.

We will give some freedom to the fitted parameters to take values withing a certain physically reasonable range: $A,B  \in (0, 50)$ kpc, $C \in (0, z_{max})$  kpc, $\omega \in (0, 4\pi)$ rad/Gy, $\phi \in (0, 2 \pi)$ rad and $\textbf{r}_0 \in (-50, 50)$   kpc. To fit the parameters we follow a squared-distance minimization strategy, in which we minimize the sum running over each of the stars in the cloud $\chi^2 (A, B, C, \omega, \phi, \textbf{r}_0 ) = \sum^N_{i=1} (\textbf{r}_i-\textbf{r}(t_i))^2, $ in this equation we compare $ \textbf{r}_i$, which are the ``observed" positions of the point-stars (simulation or data), and $\textbf{r}(t_i)$, which are the ``theoretical" positions of these point-stars at a certain time calculated for the different gravitational source shapes.

\section{\label{sec:level2} Torsion: A new mathematical tool to study DM}

In this work we represent stellar streams as parametric curves  $\textbf{r}(t)$  wrapping around galaxies and deploy an invariant property of curves, the torsion $\tau$, to probe the existent nonsphericity of a DM halo surrounding galaxies. 
In differential geometry we can parametrize curves $\textbf{r}(t)$ by an arbitrary variable $t$ that can be traded for the arclength  $s(t) = \int  |\textbf{r}(t)'| dt$ \cite{2017Carmo}. At each point $P$ in the curve there is a trihedron formed by the tangent vector $\textbf{T}=d\textbf{r}/ds$, the normal vector $\textbf{N}=d\textbf{T}/ds$ and the binormal perpendicular to both, $\textbf{B}$. 
Given a curve $\textbf{r}(t)$ the torsion measures how sharply it is twisting out of the osculating plane, instantaneously defined by the velocity and
normal acceleration. The torsion is defined as the variation of the binormal vector $\textbf{B}=\textbf{T}\times\textbf{N}$, following expression $\tau = -\frac{d \textbf{B}}{ds}\cdot \textbf{N}$ and we can also find an expression for the torsion when the curve is described by an arbitrary parameter, t, with $\tau=\frac{({\bf r}'\times {\bf r}'')\cdot {\bf r}'''}{|{\bf r}'\times {\bf r}''|^2} $.

When a curve lies on a plane, this is defined by the tangent and normal vectors and then the binormal vector will experience no variation giving a null torsion, $\tau = 0$. But when the curve twists, the situation is different because there is a variation of the binormal vector. For example, in the ideal case of a circular helicoidal movement the torsion is constant $\tau=C$. 
We will consider different shapes for gravitational sources and obtain the expected torsion using Euler's Method.\\

\begin{figure}
    \includegraphics[width=.25\linewidth]{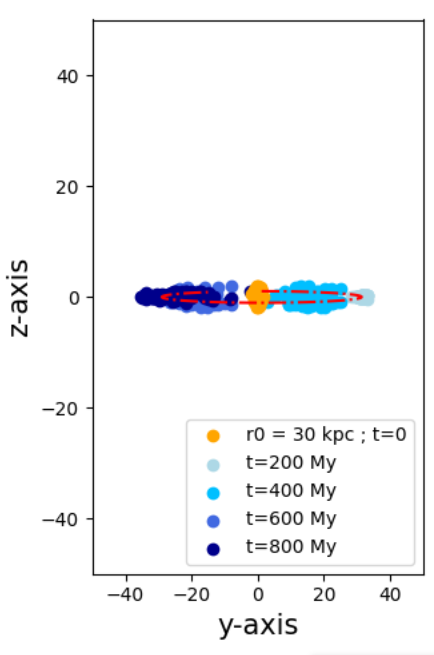}    
    \includegraphics[width=.25\linewidth]{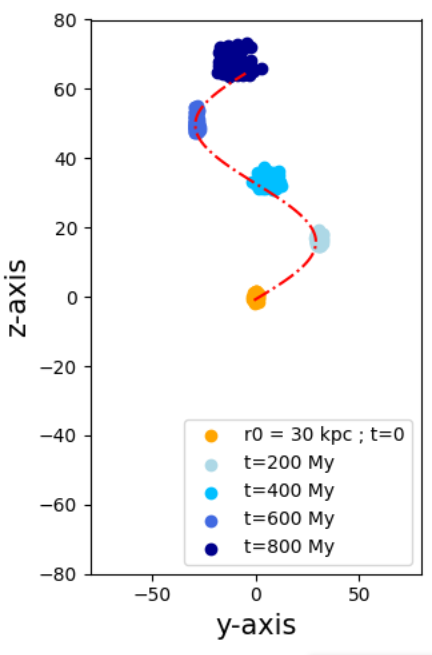}    
    \includegraphics[width=.25\linewidth]{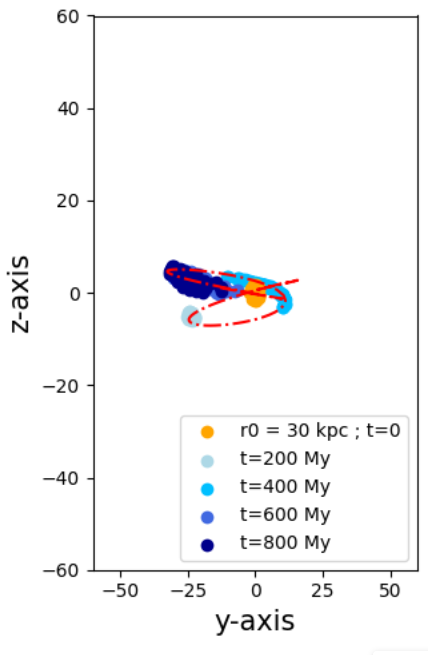}
    \caption{\label{fig:simulations} Simulated movement of a 100-star cluster around spherical (left), cylindrical (center) and combined (right) sources.  }
\end{figure}

\textbf{Planar orbits around a spherical source.}
The orbit of a massive test body around a central-field, spherically shaped gravitational source, is planar and has a null torsion (unless there are additional external forces). A cluster made of test bodies moving around a spherical galaxy halo will lose dust grains forming a kind of contrail, but its shape through space will be a planar curve. 
We can quantify the value of the torsion as function of time for the simulation in Fig.~(\ref{fig:simulations}). There we apply an analytical helicoidal fit to the resulting stream with $\chi^2/N_{gdl}=3.83$, which yields a value of $\tau= 1.5 \cdot 10^{-4}$ $kpc^{-1}$. This must be considered our level of numerical noise due to the simulation and fitting procedures and provides a floor under which actual torsion cannot be measured in a realistic system.\\

\textbf{Helicoidal orbits around a cylindric source.}
For the prior spherical field, if we were to add an initial vertical velocity to the cluster, the final orbit would still be planar --but tilted. Simulating instead a cylindrical source, an initial vertical velocity provides a helicoidal movement wrapping around the z axis, but a purely azimuthal one a planar curve. 
The extracted value for the torsion in the simulation in Fig.~(\ref{fig:simulations}) yields 
 $\tau=1.0\cdot 10^{-2}$ $kpc^{-1}$ from a helicoidal fit of the stream with a  $\chi^2/N_{gdl}=0.95$.
We could also consider the option of a rather ellipsoidal shape, the cylinder being the extreme toy-model example, and also the combination of different kinds of sources such as a spherical galaxy with an ellipsoidal contribution of a DM halo. This would lead to stellar streams with a non-zero, non-constant torsion, see Fig.~(\ref{fig:simulations}) for different times.\\

\textbf{Galactic background torsion.}
The galaxy itself, due to its internal aspherical structure, can have a non-negligible stellar-stream torsion. We wish to have a reference for a minimum torsion that could be considered ``normal'' in order of magnitude, so that if extensive studies of stellar streams show that their torsion exceeds that level, one could reject the hypothesis of a spherical halo. When we consider a galaxy as composed by a disk- plus a sphere-shaped distributions of matter, we find a background value of the order $\tau\sim O(10^{-4})$ $kpc^{-1}$ (see \cite{bariegoquintana2024torsion} for full computations). 

Torsion has dimensions of inverse length, so from the equation for the torsion we can expect stellar streams to show an inverse relation of $\tau$ with their distance from the galactic center. In a galaxy such as the Milky Way the torsion of galactic streams should have a characteristic scale of $(10$ kPc$)^{-1}$. 

\section{\label{sec:level4} Real case scenario: \ \  
The Milky Way Streams }

Since we are interested in the analysis of stellar streams wrapping around galaxies to extract information on the shape of dark matter haloes surrounding the galaxies, we can best begin with our own, the Milky Way (MW). The MW stellar streams have been investigated for some time and are still an active field of research. We will use them to infer the shape of the DM halo of our galaxy. For that, we will use galstreams as a tool developed to fit some of the best known streams in the MW \cite{10.1093/mnras/stad321}.  

This database includes many streams, but we will only consider those that are 
more sensitive to a large part of the halo, namely stellar streams at distances farther than $30$ kpc: Cetus-Palca, Cetus, Elqui, Eridanus, Jet, Pal15, Palca, Sagittarius, Willka-Yaku, Orphan-Chenab and Styx. We must discard the last two because they may be influenced by gravity sources that are outside the MW such as the Large Magellanic Cloud \cite{Conroy_2021}.

We fit the streams of the database using an helicoidal parametrization with an arbitrary parameter, t, that takes values in the interval $t \in (0, 1)$.  
The torsions for each of the streams are compiled in Table~1 in \cite{bariegoquintana2024torsion}. The local torsion along many of the streams takes significant values, above the expected $O(10^{-4})$ ``noise'' that we found in the simulations when the streams evolved around aspherical sources, as well as the $O(10^{-3})$ floor from including the galactic plane. Higher values for the torsion have been found for streams such as Cetus, Willka-Yaku, Cetus-Palca and Sagittarius than for the rest. 

We expect their respective torsions to show an inverse relation with their distance from the galactic center. We selected the streams at $30$ $kpc$ or more, meaning that we would consider values of the torsion  $\tau \sim 0.01$ $kpc^{-1}$ to be different from zero, and Table~1 in \cite{bariegoquintana2024torsion} shows several such. Also, all those with $\tau>$ 0.001 $kpc{-1}$ could perhaps carry interesting information about the DM distribution causing a serious discrepancy with the hypothesis that the DM halo is purely spherical. 

\section{\label{sec:level5} Conclusion }
The problem of galactic rotation curves suggests the presence of substantial amounts of DM surrounding galaxies, yet the precise arrangement of these sources remains unknown. While spherical DM distributions around galaxies require precise adjustments to account for the flatness of rotation curves, a cylindrical or prolate DM source can naturally explain this flattening without the need for fine-tuning. However, observations within the galactic plane are unable to distinguish between spherical haloes with isothermal profiles and cylindrical or elongated gravitational sources with arbitrary density profiles. Yet, information beyond the galactic plane could yield new discriminants.

Stellar streams can be characterized by their torsion. Around a central potential, orbits are contained within a plane and are thus expected to be torsionless, see Fig.~\ref{fig:simulations}. Conversely, test masses around cylindrical sources are expected to follow helicoidal orbits with nonzero torsion if given both initial vertical and azimuthal velocities. Another natural geometry is an ellipsoid-shaped halo, which deviates from perfect cylindrical symmetry yet exhibits elongation. The expected orbits of stream components would result from a combination of orbits around central potentials and cylinders, as seen in Fig.~(\ref{fig:simulations}). Torsion therefore serves as an observable specifically customized to assess the prolateness of a DM halo.

From our evaluation of the torsion data from the Milky Way streams, it becomes evident that it is non-negligible in certain considered streams. We do not dare favor one or another interpretation of the DM halo shape in the view of current data. However, we do observe streams exhibiting significant torsion, which is promising and indicates that further investigation could potentially shed light on the shape of the halo.

\nocite{*}
\bibliography{aipsamp}

\end{document}